\documentstyle[aps,prl,multicol,epsf]{revtex}
\begin{document}
\draft
\title{
Noise-assisted mound coarsening in epitaxial growth
}
\author{Lei-Han Tang$^{(1,3)}$, P. \v{S}milauer$^{(1,2)}$, and 
D. D. Vvedensky$^{(1)}$}
\address{
$^{(1)}$ The Blackett Laboratory, Imperial College,
London SW7 2BZ, United Kingdom\\
$^{(2)}$ Institute of Physics, Cukrovarnick\'{a} 10, 162~00~Praha~6, 
Czech Republic\\
$^{(3)}$ Department of Physics, Hong Kong Baptist University,
Kowloon Tong, Hong Kong
}

\date{\today}
\maketitle
\begin{abstract}
We propose deposition noise to be an important factor
in unstable epitaxial growth of thin films.
Our analysis yields a geometrical relation $H=(RWL)^2$
between the typical mound height $W$, mound size $L$, and 
the film thickness $H$.
Simulations of realistic systems
show that the parameter $R$ is a characteristic of the growth conditions,
and generally lies in the range $0.2$-$0.7$.
The constancy of $R$ in late-stage coarsening yields
a scaling relation between the coarsening exponent $1/z$ and the mound
height exponent $\beta$ which, in the case of
saturated mound slope, gives $\beta=1/z=1/4$.
\end{abstract}

\begin{multicols}{2}

One of the currently-contemplated applications of
vapor phase epitaxy is the fabrication of nanoscale quantum dots or wires.
Under suitable conditions, pyramids or mounds
form spontaneously on the film surface as a result of 
unstable growth. From the device point of view,
the basic challenges lie in one's ability to (i) control the size and 
shape of individual mounds and (ii) enforce great regularity
and uniformity in the mound array.
To achieve these ends, an understanding of the relation between the
growth morphology and the underlying atomic processes 
under nonequilibrium growth conditions is desirable.

Moderate surface modulations, as opposed to isolated three-dimensional
islands, are observed when there is either a small lattice mismatch
between the film and substrate, or a surface
diffusion barrier at step edges, known as the Ehrlich-Schwoebel (ES)
barrier\cite{es}. The second mechanism, which is the focus of this paper,
is purely kinetic in origin and thus operates also in homoepitaxy.
Villain\cite{villain} pointed out that presence of the ES barrier
leads to a nonequilibrium uphill surface mass current which, in the case
of a low Miller index surface, amplifies weak height fluctuations
during growth. 
Atomistic and continuum models which incorporate this effect
indeed develop the growth instability\cite{johnson,siegert,pavel,krug}.
Numerical simulations have shown that, after an initial transient
period, mounds of finite slope appear at the film surface,
as illustrated in Fig. 1.
The pattern of mounds is surprisingly regular, with a characteristic
mound size $L$ which coarsens with the film thickness $H$ as
\begin{equation}
L\sim H^{1/z}.
\label{size-scaling}
\end{equation}
The exponent $1/z$ generally lies in the range $0.15$-$0.25$.
The typical height of the mounds $W$
can also be fitted to a power-law,
\begin{equation}
W\sim H^\beta,
\label{mound-height}
\end{equation}
where the exponent $\beta$ can be as small as $0.25$, or as big as $0.5$.
Both the mound patterns seen in simulations and the range of exponent values
correlate well with some experimental findings\cite{thurmer,stroscio},
but the theoretical understanding of
the coarsening characteristics is still incomplete.
\begin{figure}
\epsfxsize=8truecm
\epsfbox{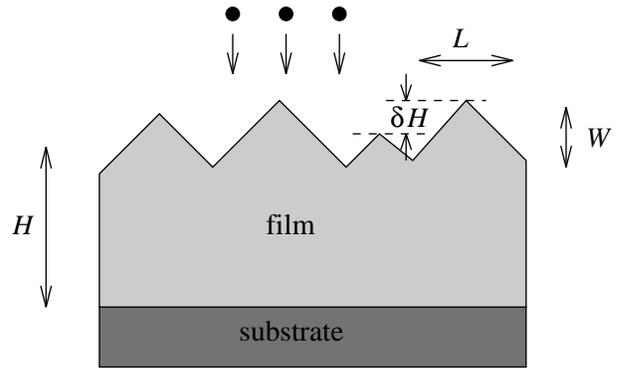}
\smallskip
\narrowtext
\caption{Schematic illustration of the mound morphology
during epitaxial growth. The typical mound size $L$ 
increases with the film thickness $H$.}
\end{figure}

Previous analytical studies of mound coarsening 
have focused on deterministic, nonlinear evolution equations derived from
a phenomenological, slope dependent surface current
\cite{johnson,siegert,kps,politi,rost,golub}.
Analogies have been made to domain coarsening in magnetic
systems, which is itself a difficult and open problem.
In this Letter we suggest an alternative scenario of mound coarsening
assisted by noise in the deposition flux.
We describe some of the measurable consequences based on this 
kinetic pathway, and discuss the interplay between noise-driven
fluctuations and deterministic surface evolution.
It is hoped that these considerations will lead to
a quantitative scheme for assessing the effect of growth conditions 
on mound morphology in both atomistic modelling and experiments.

The mound morphology illustrated in Fig. 1 is usually characterized
by two parameters: the lateral mound size $L$ and the typical 
mound height $W$. (In this paper we focus on the isotropic case,
where mounds typically have a nearly circular shape.)
For the purpose of the following discussion, we introduce a
third parameter, the typical height difference between neighboring mounds
$\delta H$. To see if the deposition noise is a significant factor
in coarsening, let us first consider a simplified model, assuming
(i) $\delta H$ is solely due to fluctuations
in the local deposition rate,
and (ii) $\delta H$ is a fixed fraction of $W$.
(The justification of these assumptions will be considered below.)
After $H$ layers of material have been deposited, 
the number of deposited atoms in the column under a given mound
has a fluctuation
$\delta N\simeq (HL^d)^{1/2}$,
where both $L$ and $H$ are measured in units of atomic spacing,
and $d$ is the dimensionality of the surface.
This yields the estimate
\begin{equation}
\delta H\simeq \delta N/L^d=(H/L^d)^{1/2}. 
\label{height-flu}
\end{equation}
{}From assumption (ii), we obtain,
\begin{equation}
H\simeq W^2L^d.
\label{noise-scaling}
\end{equation}
In the regime where the power-laws (\ref{size-scaling})
and (\ref{mound-height}) are well obeyed, Eq. (\ref{noise-scaling}) yields,
\begin{equation}
{2\beta\over d}+{1\over z}={1\over d}.
\label{scaling-law}
\end{equation}
If the mound slope $s\simeq W/L$ saturates to a constant,
the above equation gives,
\begin{equation}
\beta={1\over z}={1\over d+2}.
\label{beta}
\end{equation}

Kawakatsu and Munakata\cite{km} carried out a detailed study of a
one-dimensional noise-driven coarsening model which 
confirms the scaling result derived above.
In that model, the slope $s$ saturates to a finite value after an
initial transient. From Eq. (\ref{beta}) one obtains $z=3$
which agrees with their analysis.
In comparison, the noiseless model yields a much slower
(logarithmic) coarsening law for $d=1$ \cite{politi}.

In the physically relevant case $d=2$,
we have checked that Eq. (\ref{scaling-law}) is consistent
with existing numerical studies.
For models which exhibit a saturated
mound slope, several groups have concluded that $\beta=1/z=1/4$ 
\cite{siegert,krug}, as suggested by (\ref{beta}).
Higher values\cite{pavel} of $\beta$ 
(and hence lower values of $1/z$)
are obtained if the slope $s$ continues to increase with
the film thickness $H$, in agreement with Eq. (\ref{scaling-law}).

The apparent success of the simplified model 
points to the relevance of the deposition noise in the coarsening process,
contrary to the prevailing belief in the literature. 
To appreciate the significance of the noise on a more quantitative level, 
one needs to examine various surface mass transport processes,
and see how they might modify the assumptions of the simplified model.
In general terms, surface mass transport is governed by bonding 
energies of atoms at steps, and by various activation energies
for hopping on the terrace, along a step (or ledge), and up or
down a step, etc. The resulting surface mass current has a
deterministic component whose direction is mainly determined 
by the distribution of bonding sites and their strengths, 
often modelled by a chemical potential field, and a stochastic 
component due to the intrinsic random nature of the hopping process.
It is understood that the substrate temperature has a dramatic
influence on the kinetic coefficient which controls the
magnitude of the current.
In what follows we shall only consider the deterministic component
of the surface current.

The surface dynamics described above plays a dominant role
in the initial development of the growth instability, and also
sets the saturated (or quasi-stationary) slope of the mounds
in the late-stage coarsening process. However, the ability
of the surface dynamics in promoting mass transport is greatly
reduced after quasi-stationary mounds are well-developed.
This is because the very shape of the mounds is chosen to
minimize the surface current either uphill or downhill\cite{kps}.
In this regime, each mound has acquired a form of metastability
regarding its shape. Fluctuations in the amount of deposited material 
can be incorporated by enlarging the overall size of the mound,
with minimal amount of inter-mound mass transport.
In such a situation, the mounds can be treated as independent fluctuating
entities, as in the simplified model.
The height difference between neighboring mounds, however,
is subject to the geometrical constraint $\delta H\leq W$.
In the absence of other requirements, we arrive at assumption (ii)
of the simplified model.

To elaborate on this view, we estimate now the magnitude
of the deterministic mass transport between two neighboring mounds 
due to the surface dynamics for the geometry illustrated in Fig. 2.
The center of each mound consists of roughly concentric rings of steps,
and the distance between the two centers is denoted by $L$.
The two mounds are joined by a ``ridge terrace''. The outer rim
of the ridge terrace has convex parts on either side
and concave parts in the middle. Sites on the concave parts
on average offer better lateral bonding, and hence are
energetically more favorable. This effect can be modelled
by a chemical potential difference $\delta\mu$ between
the convex and concave parts, which should be proportional to
the curvature, i.e., $\Delta\mu\sim 1/L$.
An inward mass current is thus expected to appear,
as shown in Fig. 2, with a magnitude 
\begin{equation}
j_s\simeq D_s\Delta\mu/L,
\label{j_s}
\end{equation}
where $D_s$ is a kinetic transport coefficient.
If we ignore interlayer mass transport,
which is a reasonable assumption given the instability,
the kinetic pathways which contribute to $j_s$
are then (a) diffusion along the step,
and (b) detachment of atoms from the step to the terrace, which
in this case is a narrow strip one layer below the ridge terrace,
followed by diffusion on the terrace. In both cases the transport
is essentially a one-dimensional process so $D_s$ does not
depend appreciably on $L$, but it may depend on the strip
width (which is inversely proportional to the mound slope $s$)
if (b) dominates.
The same mechanism is expected to operate also on other layers below
the ridge terrace, though $j_s$ decreases due to decreasing curvature.
Assuming the effect extends to $Q$ layers, the total inward mass
current is given by,
\begin{equation}
J_s\simeq Qj_s\simeq D_s\Delta\mu{Q\over L}.
\label{total-current}
\end{equation}

\begin{figure}
\epsfxsize=8truecm
\epsfbox{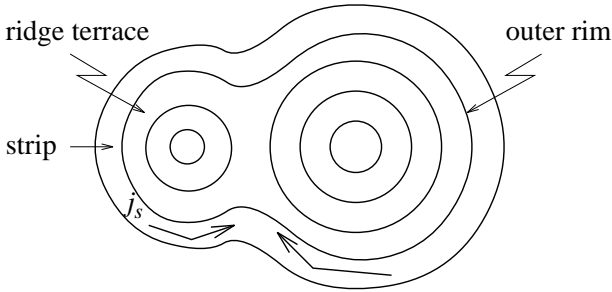}
\smallskip
\narrowtext
\caption{Top view of two neighboring mounds of unequal size.
Better lateral bonding for surface atoms is achieved at the
concave parts of the closed steps.
This mechanism results in an inward mass current $j_s$.
}
\end{figure}

If we compare the shape of the ridge terrace, each time after
exactly one monolayer of atoms is deposited,
we are likely to see a rounding of the ridge terrace
due to the current $j_s$. From the geometry, it is
seen that rounding encourages the upper rings of the 
two mounds to move towards each other, as the inner part
of the ridge terrace receives increasingly more flux of atoms from
the beam. This eventually drives the two mounds to merge with
each other. Although details of the coarsening process are
likely to be complicated, an estimate of the time scale $\tau_s$
can be obtained by equating the total
mass $M$ carried by $J_s$ to the volume of the gap between the two mounds,
i.e., $M\simeq L^2W$. This yields a time scale for coarsening
due to surface dynamics,
\begin{equation}
\tau_s=M/J_s\simeq L^4{W\over D_s Q}.
\label{tau-s}
\end{equation}

It is interesting to see that, taking $Q\simeq W$, Eq. 
(\ref{tau-s}) also yields $z=4$, suggesting that surface transport
driven by bonding energies may play an equally important role
in the coarsening process.
One should note, however, that the effectiveness of this mechanism
is governed by the kinetic coefficient
$D_s$, which is a strong function of the substrate temperature.
In addition, in the presence of other mounds, there are
competing tendencies for the surface mass flow, resulting in
an increase of $\tau_s$ and further limiting the effectiveness
of surface dynamics. It is natural to expect the
latter effect to be particularly significant when mounds are 
of nearly equal size. 

Our estimate of the coarsening time (\ref{tau-s}) based
on surface dynamics alone is consistent with the results
in a recent paper by Rost and Krug\cite{rost}, who gave an upper bound $1/4$
for the coarsening exponent $1/z$ based on the analysis
of a deterministic continuum model.
Hence for the case $d=2$, the timescale set by the deterministic 
surface dynamics is slow enough to allow a role for deposition noise 
in the actual coarsening process.

A plausible scenario in the noisy case can thus be stated as follows.
In the late-stage coarsening regime where mounds have acquired
their quasi-stationary shape, deposition noise is responsible
for the height difference (or equivalently, size difference) 
between neighboring mounds when this difference is small.
In this case, the height of each mound fluctuates independently
(i.e., each mound dances up and down in a random fashion).
A ``cap value'' (or threshold) exists on the height difference,
either due to the geometry (as assumed in the simplified model),
or due to a crossover to a relatively rapid, surface-dynamics-driven merging
when the disparity in mound size becomes too big to sustain.

As a quantitative measure of the cap value, we introduce the ratio,
\begin{equation}
R={\delta H\over W}={H^{1/2}\over WL^{d/2}}.
\label{ratio}
\end{equation}
It is easy to see that $R$ also measures the ratio between the
excess material in a given mound due to fluctuations in the
deposition rate, $\delta N$, and the total mound volume $WL^d$.
A constant $R$ during growth can be interpreted as supporting
the noise-assisted coarsening picture proposed above, while
a decreasing $R$ with $H$ would suggest a diminishing role of
deposition noise in late-stage coarsening.

The geometrical parameters $L$ and $W$ in Eq. (\ref{ratio})
can be defined precisely
through the height correlation function \cite{stroscio},
\begin{equation}
G({\bf x},t)=\langle h({\bf x}_0,t) h({\bf x}_0+{\bf x},t)\rangle,
\label{two-pt}
\end{equation}
where $h({\bf x},t)$ is the height fluctuation at point ${\bf x}$ 
on the surface at time $t$.
The width of the surface $W$ can be identified with the
root-mean-square deviation of $h$, i.e., $W=G^{1/2}(0,t)$.
For a mounded surface, $h$ becomes anti-correlated over
a distance of mound size. We can thus identify $L$ with the
mound radius, i.e., where the spherically averaged 
$G({\bf x},t)$ reaches its first zero starting from the origin. 
[For an isotropic surface
$G({\bf x},t)=G(|{\bf x}|,t)$, and hence $L$ satisfies $G(L,t)=0$.]

Using the above definition, we have computed $R$ from both 
published\cite{pavel} and unpublished\cite{pavel2} simulations.
Figure 3 shows part of our data for two sets of growth parameters
at various temperatures. 
The first set, with growth parameters simulating homoepitaxy of 
GaAs(001) at a deposition rate $F={1\over 6}$ monolayers (ML)/s, 
are for substrate temperatures $T=678$~K (filled diamond),
$778$~K (filled square), and $828$~K (filled circle).
The typical mound slope in this case is of the order of
0.05, and the mound size can be as big as
50 lattice constants after depositing 1000 ML's of atoms.
In the second set,
the growth parameters were chosen to model Pt(111) homoepitaxy
at a deposition rate $F={1\over 40}$ML/s, and substrate temperatures
$T=350$~K (open diamond) and $400$~K (open circle).
Here the mounds are typically much smaller, but the slopes are
much higher, in accordance with Eq. (\ref{noise-scaling}).
Apart from statistical fluctuations,
it is seen that in each case the parameter $R$ reaches
a constant after an initial transient. The value of
$R$ falls in the range $0.2$--$0.7$.
Within a given set of surface energy parameters, we see that
$R$ decreases with increasing temperature, in agreement
with our expectation that surface dynamics plays a more
important role at higher temperatures due to a much larger
kinetic coefficient $D_s$.
We have also examined the data with the same surface activation
energies as those shown but different values for the ES barrier,
or a different transient dynamics for an impinging adatom.
It is found that the latter parameters, although have very strong 
effects on the mound slope, do not change the $R$ value as 
significantly as the growth temperature $T$.
Details of the analysis will be published elsewhere \cite{pavel2}.

\begin{figure}
\epsfxsize=8truecm
\epsfbox{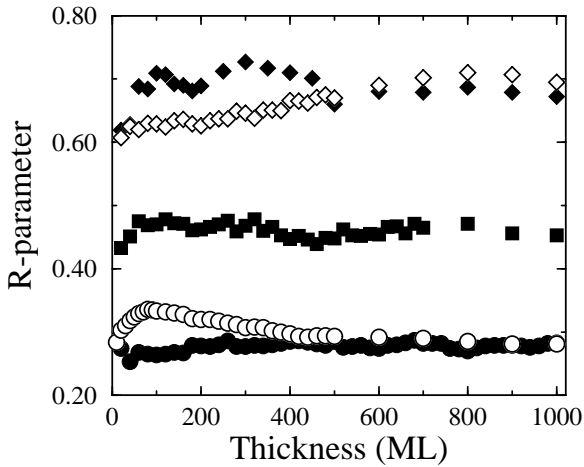}
\smallskip
\narrowtext
\caption{Simulation data showing the constancy of the parameter $R$ 
during growth. Filled symbols correspond to 
GaAs(001) surface at a deposition rate $1/6$ ML/s and
substrate temperatures
$T=678$~K (diamond), $778$~K (square), and $828$~K (circle).
Open symbols correspond to Pt(111) surface
at a deposition rate $1/40$ ML/s and substrate temperatures
$T=350$~K (diamond) and $400$~K (circle).}
\end{figure}

The main conclusion of our study is that deposition noise is an 
important factor in driving mound coarsening after
the initial growth instability due to the ES barrier is
well-developed. The picture we developed leads to an important geometric
relation between the
parameters $L$ and $W$ characterizing the mound morphology,
and the film thickness $H$, with a proportionality constant $R$
of order unity when these parameters are measured in atomic
units. The constancy of $R$ reveals a scaling relation between
the mound coarsening exponent $1/z$ and the mound height
exponent $\beta$, which agrees with previous simulation results.
Simulations of realistic growth conditions indeed show the 
constancy of $R$ during growth, but the saturated value
decreases with increasing substrate temperature. This 
is supported by our semi-quantitative
analysis of surface mass transport driven by bonding energies.

It is difficult to over-emphasize the utility of Eq. (\ref{ratio})
in experiments where one wishes to predict quantitatively
the mound size as a function of the film thickness.
Knowing the mound slope and the coefficient $R$, which
is a function of the growth conditions only, it is possible
to calculate $L$ for a given film thickness $H$.
Our preliminary investigation shows that the most significant
influence on the value of $R$ comes from the substrate temperature.
In addition, a smaller value of $R$ indicates a weaker variation
in the height or size of the mounds, which may be desirable
for certain optical device applications.

P.\v{S}. acknowledges financial support of the Grant No. 202/96/1736 
of the Grant Agency of the Czech Republic.

\end{multicols}

\end{document}